\documentclass[a4paper]{jpconf}
\usepackage{graphicx}
\usepackage{amsmath,amssymb}
\usepackage{mathabx}
\usepackage{tipa,tipx}
\usepackage{epstopdf}

\begin{document}
\title{The Effect of Different Magnetospheric Structures on Predictions of Gamma-ray Pulsar Light Curves}

\author{M Breed$^1$, C Venter$^1$, A K Harding$^2$, T J Johnson$^3$}

\address{$^1$Centre for Space Research, North-West University, Potchefstroom Campus, Private Bag X6001, Potchefstroom, 2520, South-Africa} 
\address{$^2$Astrophysics Science Division, NASA Goddard Space Flight Center, Greenbelt, MD 20771, USA} 
\address{$^3$NRC Fellow, High-Energy Space Environment Branch, Naval Research Laboratory}

\ead{20574266@nwu.ac.za}

\begin{abstract}
The second pulsar catalogue of the {\it Fermi} Large Area Telescope (LAT) will contain in excess of 100 gamma-ray pulsars. The light curves (LCs) of these pulsars exhibit a variety of shapes, and also different relative phase lags with respect to their radio pulses, hinting at distinct underlying emission properties (e.g., inclination and observer angles) for the individual pulsars. Detailed geometric modelling of the radio and gamma-ray LCs may provide constraints on the B-field structure and emission geometry. We used different B-field solutions, including the static vacuum dipole and the retarded vacuum dipole, in conjunction with an existing geometric modelling code, and constructed radiation sky maps and LCs for several different pulsar parameters. Standard emission geometries were assumed, namely the two-pole caustic (TPC) and outer gap (OG) models. The sky maps and LCs of the various B-field and radiation model combinations were compared to study their effect on the resulting LCs. As an application, we compared our model LCs with {\it Fermi} LAT data for the Vela pulsar, and inferred the most probable configuration in this case, thereby constraining Vela's high-altitude magnetic structure and system geometry.
\end{abstract}

\section{Introduction}

Pulsars are considered to be cosmic lighthouses that rotate at tremendous rates and are highly magnetized neutron stars (NS) \cite{Chaisson}. The fact that pulsars are embedded in such extreme conditions make them valuable laboratories for studying a wide range of topics, including: nuclear physics, plasma physics, electrodynamics, magnetohydrodynamics (MHD), and general relativistic physics \cite{Becker}. Pulsars emit radiation across the electromagnetic spectrum, including radio, optical, X-ray and gamma ($\gamma$) rays \cite{Chaisson}. We focus on $\gamma$-ray pulsars, specifically the Vela pulsar, which was detected \cite{Abdo} by the {\it Fermi} Large Area Telescope (LAT) \cite{Atwood}. The Vela pulsar is the brightest persistent GeV source. {\it Fermi} was launched in 2008 and has discovered in excess of 100 $\gamma$-ray pulsars. The LAT has a very large field of view of 2.4 sr which enables it to observe 20\% of the sky at any instant, scanning the entire sky in a time frame of a few hours. 

\subsection{Geometric pulsar models} 

There are several models which can be used for the modelling of high-energy (HE) emission from pulsars. These models include the two-pole caustic (TPC) (the slot gap (SG) \cite{Muslimov} model may be its physical representation) model \cite{Dyks}, outer gap (OG) model \cite{Romani},\cite{Cheng} and pair-starved polar cap (PSPC) model \cite{Harding}. Figure 1 illustrates these geometric pulsar models, and their emission regions \cite{Dyks}.

\begin{figure}[t]
\begin{center}
\includegraphics[width=9cm]{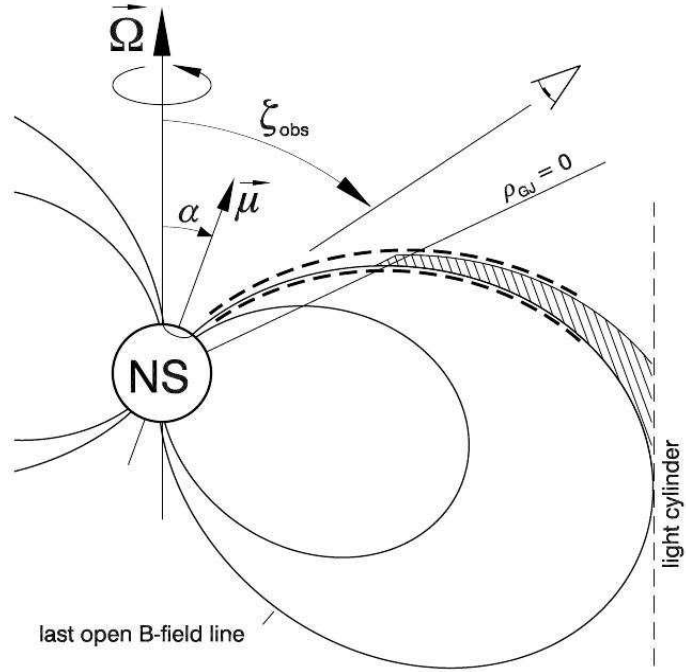}
\end{center}
\caption{\label{geom} A schematic representation of geometric pulsar models. The TPC emission region extends from $R_{\rm{NS}}$ (NS radius) up to $R_{\rm{LC}}$ (light cylinder radius), OG region from $R_{\rm{NCS}}$ (null charge surface radius) to $R_{\rm{LC}}$, and the PSPC from $R_{\rm{NS}}$ to $R_{\rm{LC}}$ [covering the full open volume region].}
\end{figure}   
  
Consider an ($\vec{\Omega}$, $\vec{\mu}$) plane, with $\vec{\mu}$ (the magnetic moment) inclined by an angle $\alpha$ with respect to the rotation axis $\vec{\Omega}$ (the angular velocity). The observer's viewing angle $\zeta$ is the angle between the observer's line of sight and the rotation axis. The gap region is defined as the region where the relativistic particles originate and particle acceleration takes place. The emissivity of HE photons within this gap region is assumed to be uniform in the co-rotating frame and the $\gamma$-rays are expected to be emitted tangentially to the local magnetic field in this frame \cite{Dyks_a}. The gap region for the TPC model extends from the surface of the NS along the entire length of the last closed magnetic field lines, up to the light cylinder, as indicated by the dashed lines in Figure~\ref{geom}. For the OG model, the gap region extends from the null-charge surface (NCS), where the Goldreich-Julian charge density is $\rho_{GJ}=0$ \cite{GJ}, to the light cylinder, as indicated by the shaded region (the emission region may be located at slightly smaller co-latitudes compared to the TPC). The PSPC is the gap region that extends from the surface of the NS to the light cylinder over the full open volume \cite{Harding}. 

\subsection{Assumed magnetospheric structure}

The magnetospheric structures studied in this paper include the static \cite{Griffiths} and retarded vacuum dipole \cite{Deutsch}. The (aligned) static dipole is a special case of the retarded dipole and is described by the following B-field equations in terms of spherical coordinates in the laboratory frame:

\begin{equation}
B_{\rm st,r} = \frac{2\mu}{r^3}\cos\theta
\end{equation}

\begin{equation}
B_{\rm st,\rm \theta} = \frac{\mu}{r^3}\sin\theta.
\end{equation}

The retarded dipole is described by the following B-field equations \cite{Dyks_b}: 

\begin{equation}
B_{\rm ret,\rm r} = \frac{2\mu}{r^3}[\cos\alpha\cos\theta  + \sin\alpha\sin\theta(r_{\rm n}\sin\lambda + \cos\lambda)]
\end{equation}

\begin{equation}
B_{\rm ret,\rm \phi} = -\frac{\mu}{r^3}\sin\alpha[(r_{\rm n}^2 - 1)\sin\lambda + r_{\rm n}\cos\lambda]
\end{equation}

\begin{equation}
B_{\rm ret,\rm \theta} = \frac{\mu}{r^3}{(\cos\alpha\sin\theta  + \sin\alpha\cos\theta[-r_{\rm n}\sin\lambda + (r_{\rm n}^2 - 1)\cos\lambda])}
\end{equation}

\begin{equation}
\lambda = r_{\rm n} + \phi - \Omega t 
\end{equation}

\begin{equation}
r_{\rm n} = \frac{r}{R_{\rm{LC}}}.
\end{equation}

By setting $r_{\rm n}$ equal to zero the retarded field simplifies to the general (non-aligned) static dipole, with {\it r} the radial distance. The static dipole field is studied for numerous reasons. Two of them are: (1) calculations are simpler for this B-field, and (2) when the results for the static dipole are compared to those for the other B-fields, the importance of the near-$R_{\rm{LC}}$ distortions in the B-fields for predicted radiation characteristics can be gauged \cite{Dyks_c}. 

In this paper we will study the impact of different magnetospheric structures on the predictions of $\gamma$-ray pulsar LCs.
The layout is as follows: \S 2 describes the method we used to construct sky maps, LCs, and $\chi^{2}$ contour plots for the different combinations of the two B-fields and two geometric models. Section 3 contains our results, \S 4 our discussions and \S 5  contains the conclusions and future aims.

\section{Method}

\begin{table}[b]
\begin{small}
\caption{\label{ex} Best-fit ($\alpha$, $\zeta$) values for the Vela pulsar.}
\begin{center}
\begin{tabular}{lccccccc}
\br
& & \multicolumn{2}{c}{Our model} & \multicolumn{2}{c}{Reference fit~\cite{Watters}}&\multicolumn{2}{c}{Radio polarization}\cite{Johnston} \\
Combination&$\log_{10}{\chi^{2}}$&$\alpha$($^\degree$)&$\zeta$($^\degree$)& $\alpha$($^\degree$)&$\zeta$ ($^\degree$)& $\alpha$($^\degree$)&$\zeta$ ($^\degree$)  \\
\mr
{\it Static Dipole}:\\
TPC&15.3&60&85\\
OG&6.4&65&85\\
{\it Retarded Dipole}:\\
TPC&15.7&70&55&62--68&64\\
OG&1.3&80&70&75&64\\
& & & & & & 53 & 59.5\\
\br
\end{tabular}
\end{center}
\end{small}
\end{table}
We used an existing geometric modelling code \cite{Dyks_a} in which different B-field solutions and geometric models are implemented. We constructed sky maps, which are defined as the intensity per solid angle as a function of phase and $\zeta$, and LCs for the B-field and radiation model combinations, using a 5$^{\degree}$ resolution for $\alpha$ and $\zeta$. LCs are obtained by making a constant-$\zeta$ cut through each sky map. 

After the preparation of the sky maps and LCs, a statistical method for finding the best fits is applied. We used a $\chi^{2}$ method to compare our model LCs with {\it Fermi} LAT data for the Vela pulsar:

\begin{equation}
\chi^{2} = \sum_{i=1}^{\rm N}\frac{\left(Y_{\rm d,i} - Y_{\rm m,i}\right)^2}{Y_{\rm m,i}}, 
\end{equation}

with $Y_{\rm m,i}$ the model (relative flux) value and $Y_{\rm d,i}$ the measured number of counts (relative units) in each phase bin. First we lowered the model LCs resolution, so that both the model and data have the same amount of bins N. Next, we smoothed the data using a Gaussian kernel density estimator (KDE). The data are treated as being cyclic. For computational efficiency, we aligned the maximum peaks of the model and data before calculating $\chi^{2}(\alpha,\zeta)$. A contour plot of $\chi^{2}$ is shown in panel (b) of Figure~\ref{comb}. 

\section{Results}

\begin{figure}[!h]
\begin{center}
\includegraphics[width=15cm]{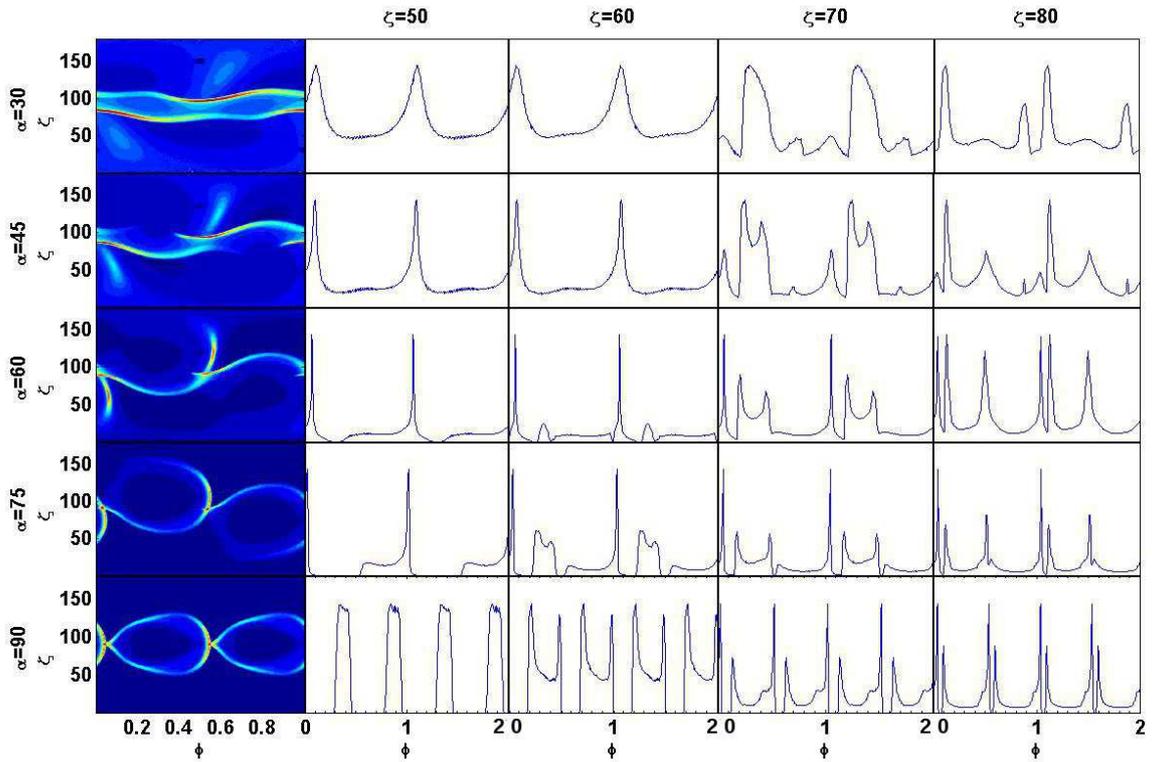}
\end{center}
\caption{\label{stat} The sky maps (left) and LCs (right) as predicted from the TPC model using the static dipole field for different $\alpha$ and $\zeta$ values (deg).}
\end{figure}   

\begin{figure}[!h]
\begin{center}
\includegraphics[width=15cm]{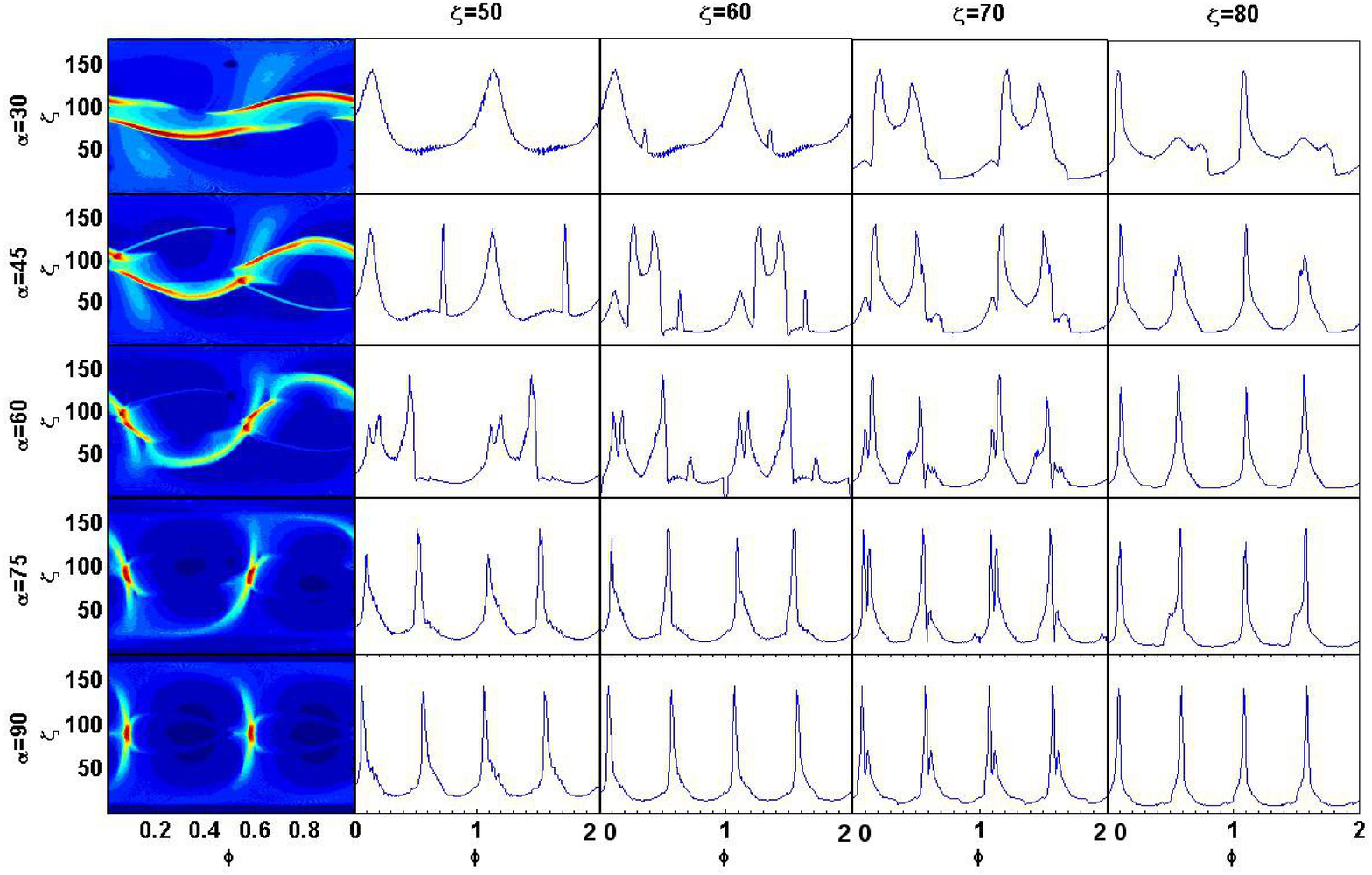}
\end{center}
\caption{\label{ret} The sky maps (left) and LCs (right) as predicted from the TPC model using the retarded dipole field for different $\alpha$ and $\zeta$ values (deg).}
\end{figure}   

\begin{figure}[!h]
\begin{center}
\includegraphics[width=15cm]{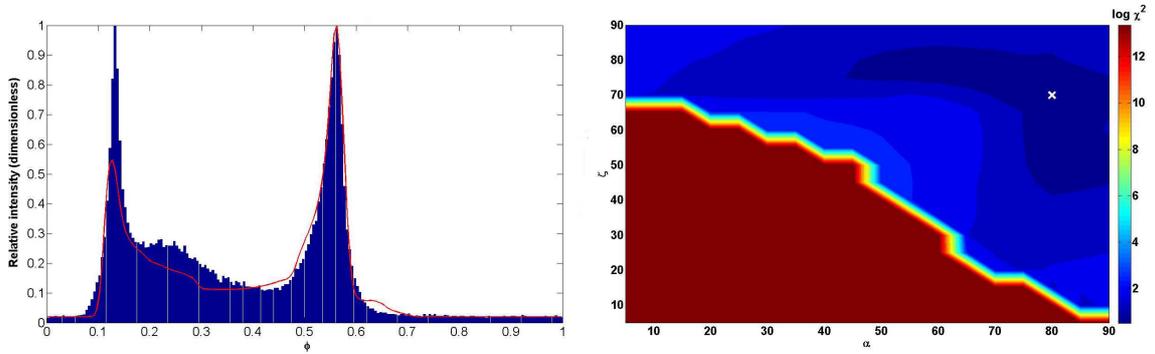}
\end{center}
\caption{\label{comb} Panel (a) indicates our best-fit LC for Vela (see Table~\ref{ex}). Panel (b) shows the contour plot for $\chi^{2}(\alpha,\zeta)$, indicating the best-fit solution.}
\end{figure} 

As an example, we show the sky maps and their corresponding LCs for the TPC model, for both the static and retarded dipole fields (Figure~\ref{stat} and \ref{ret}). We used a maximum gap radius of $R_{\rm max}=1.2R_{\rm LC}$ for both the TPC and OG cases. There are different LCs on the right of each sky map corresponding to different $\zeta$-cuts. In both the figures there appear two dark circles, the PCs, followed by two sharp, bright regions near it, called the main caustics, on the sky map. The caustic structure is qualitatively different between the two cases, leading to differences in the resulting LCs. The caustics seem wider and more pronounced in the retarded dipole case. A thin line of emission, due to the `notch' \cite{Dyks_b} is also visible in the latter case. For large $\alpha$ the caustics extend over a larger range in $\zeta$ for the retarded case compared to the static case. The OG models are not visible at all angle combinations and thus do not fill all phase space. This is due to emission that occurs below the null charge surface for the TPC model, but not for the OG model. The TPC model LCs also exhibit relatively more off-pulse emission. The LCs in the OG models are due to emission from only one pole, while both poles are visible in the TPC model. 

The different model LCs are fitted to the Vela data of {\it Fermi} LAT and for each model, we constructed a $\chi^{2}$ contour plot which indicates the best possible fit. The white marker on the contour plot (Figure~\ref{comb}, panel (b)) indicates the minimum value of $\log_{10}{\chi^{2}}$ and these values are shown in Table~\ref{ex}. The first column shows our different combinations of magnetic field and geometric model, the second indicates the minimum value of $\log_{10}{\chi^{2}}$, and the third and fourth columns indicate the best-fit $\alpha$ and $\zeta$ from our models. These are for the 5$^\degree$ resolution. We will estimate more rigorous errors on these values in future. The fifth column contains the derived $\zeta$ values from the pulsar wind nebula (PWN) torus fitting with $\alpha$ constrained \cite{Ng2008}. The last two columns are derived from fits of the rotating vector model to the radio polarization angle (PA) versus phase $\phi$ \cite{Cooke1969}. The best-fit model LC to the Vela LC is shown in Figure~\ref{comb}, panel (a). Our best fit is close to values inferred from these independent studies.       

\section{Conclusions and future work}

We have studied the effect of different magnetic fields on gamma-ray LC characteristics. We utilized the static and retarded vacuum dipole solutions, in combination with the TPC and OG geometries. It is evident that the magnetospheric structure and emission geometry determine the pulsar visibility and also the $\gamma$-ray pulse shape. We applied our models to the Vela pulsar and found a best fit from the OG model using the retarded dipole field, for $(\alpha,\zeta)=(80^\degree,70^\degree)$. This is reasonably close to the value of $(\alpha,\zeta) = (75^\degree, 64^\degree)$ inferred by \cite{Watters}. In future, we will implement an additional magnetic field solution, the offset dipole \cite{Harding_a} and study the effect of this solution on the predicted pulsar LCs.  

\section*{Acknowledgments}

This work is supported by the South African National Research Foundation (NRF). A.K.H. acknowledges the support from the NASA Astrophysics Theory Program. C.V., T.J.J., and A.K.H. acknowledge support from the {\it Fermi} Guest Investigator Program.

\end{document}